\documentclass[twocolumn,showpacs,preprintnumbers,amsmath,amssymb,superscriptaddress]{revtex4}
\usepackage{graphicx}
\usepackage{color}
\usepackage{dcolumn}
\usepackage{bm}

\begin{document}

\title{Counterflow of electrons in two isolated quantum point contacts}
\author{V.S.~Khrapai}
\affiliation{Center for NanoScience and Department f\"ur Physik,
Ludwig-Maximilians-Universit\"at, Geschwister-Scholl-Platz 1,
D-80539 M\"unchen, Germany} \affiliation{Institute of Solid State
Physics RAS, Chernogolovka, 142432, Russian Federation}
\author{S.~Ludwig}
\affiliation{Center for NanoScience and Department f\"ur Physik,
Ludwig-Maximilians-Universit\"at, Geschwister-Scholl-Platz 1,
D-80539 M\"unchen, Germany}
\author{J.P.~Kotthaus}
\affiliation{Center for NanoScience and Department f\"ur Physik,
Ludwig-Maximilians-Universit\"at, Geschwister-Scholl-Platz 1,
D-80539 M\"unchen, Germany}
\author{H.P.~Tranitz}
\affiliation{Institut f\"ur Experimentelle und Angewandte Physik,
Universit\"at Regensburg, D-93040 Regensburg, Germany}
\author{W.~Wegscheider} \affiliation{Institut
f\"ur Experimentelle und Angewandte Physik, Universit\"at
Regensburg, D-93040 Regensburg, Germany}
\begin{abstract}
We study the interaction between two adjacent but electrically
isolated quantum point contacts (QPCs). At high enough
source-drain bias on one QPC, the drive-QPC, we detect a finite
electric current in the second, unbiased, detector-QPC. The
current generated at the detector-QPC always flows in the opposite
direction than the current of the drive-QPC. The generated current
is maximal, if the detector-QPC is tuned to a transition region
between its quantized conductance plateaus and the drive-QPC is
almost pinched-off. We interpret this counterflow phenomenon in
terms of an asymmetric phonon-induced excitation of electrons in
the leads of the detector-QPC.
\end{abstract}  \maketitle

The state of a confined quantum system is modified by interactions with an
external field (or with external sources of energy). In semiconductor
nanostructures the energy and quasi-momentum of electrons acting as probe are
strongly influenced by the environment, e.\,g.\ via electron-electron or
electron-phonon interaction. If driven out of equilibrium, Coulomb forces
establish the local equilibrium  within the electron system whereas
electron-phonon interactions dominate the energy exchange with the
environment~\cite{gantmakherlevinson}. Drag experiments in semiconductor
nanostructures provide a tool to study the effect of external electrons or
phonons onto a probe electron system.

Current drag between parallel two-dimensional (2D) electron layers
has been investigated in GaAs/AlGaAs bilayer systems. At small
interlayer separations, observations are consistent with the
Coulomb drag phenomenon~\cite{gramila}. At larger separations
virtual-phonon exchange has been invoked to explain the
data~\cite{rubel95}. A negative sign of a current drag between 2D
and 3D electron gases in GaAs was explained by the Peltier
effect~\cite{solomon}. At high filling factors in a perpendicular
magnetic field a sign change of the longitudinal drag between
parallel 2D layers was found as a function of the imbalance of the
electron density  in the two layers~\cite{feng,lok}.

Interactions between two lateral quantum wires in GaAs have been
investigated in Ref.~\cite{zverev}. The observed frictional drag,
strongly oscillating as a function of the one-dimensional (1D)
subband occupation, was interpreted in terms of Coulomb
interaction between two Luttinger liquids. Recently, the
observation of negative Coulomb drag between two disordered
lateral 1D wires in GaAs in perpendicular magnetic fields was
reported~\cite{tarucha}.

Here we report on a novel interaction effect between two
neighboring quantum point contacts (QPCs), embedded in mutually
isolated electric circuits. When a strong current is flowing
through the partially transmitting drive-QPC, we detect a small
current in the second, unbiased, detector-QPC. The detector
current flows in the {\it opposite} direction of the drive current
and shows a nonlinear dependence on the source-drain bias of the
drive-QPC. It oscillates as a function of the detector-QPC
transmission. We suggest an explanation of this counterflow
phenomenon in terms of asymmetric phonon-induced excitation of
ballistic electrons in the leads of the detector-QPC.

\begin{figure}[t]\vspace{2mm}
\scalebox{0.46}{\includegraphics[clip]{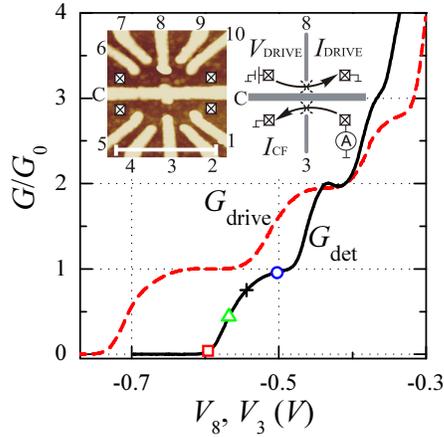}}\caption{Conductance
of the drive-QPC (dashed line) and the detector-QPC (solid line)
in the linear response regime as a function of respective gate
voltages $V_8$ and $V_3$. Symbols on the detector-QPC curve mark
the $V_3$ values used for counterflow conductance measurement
presented in Fig.~\ref{fig2}b. Left inset: AFM micrograph of the
metal gates on the surface of the heterostructure (bright tone).
Crossed squares mark contacted 2DEG regions. The scale bar equals
1~$\mu$m. Right inset: sketch of the counterflow measurement. The
directions of currents are shown for the case of
$V_\text{drive}>0$. \label{fig1}}\vspace{-0.1in}
\end{figure}

Our samples are prepared on a GaAs/AlGaAs heterostructure
containing a two-dimensional electron gas $90$~nm below the
surface, with an electron density of
$n_S=2.8\times10^{11}$~cm$^{-2}$ and a low-temperature mobility of
$\mu=1.4\times10^6$~cm$^2/$Vs. An AFM micrograph of the split-gate
nanostructure, produced with e-beam lithography, is shown in the
left inset of Fig.~\ref{fig1}. The negatively biased central gate
C divides the electron system into two separate circuits, and
prevents leakage currents between them.  Two QPCs are defined on
the upper and lower side of the central gate, respectively, by
biasing gates 8 and 3. Other gates are grounded if not stated
otherwise.

The right inset of Fig.~\ref{fig1} shows a sketch of the
counterflow experiment. We use separate electric circuits for the
(upper) drive-QPC and (lower) detector-QPC. A dc bias voltage,
$V_\text{drive}$, is applied to the left lead of the drive-QPC,
while the right lead is grounded. A current-voltage amplifier with
an input voltage-offset of about 10~$\mu$V is connected to the
right lead of the detector-QPC. Its left lead is always maintained
at the same offset potential in order to assure zero voltage drop
across the detector-QPC. In both circuits, a positive sign of the
current corresponds to electrons flowing to the left. For
differential counterflow conductance measurements, the drive bias
is modulated at a frequency of 21~Hz and the resulting ac current
component in the detector circuit is measured with lock-in
detection in the linear response regime.  All measurements are
performed in a dilution refrigerator at an electron temperature
below 150~mK. The experimental results are the same if detector
and drive QPC are interchanged.

First, we characterize the QPCs using a standard differential
conductance measurement. Figure~\ref{fig1} displays the
differential conductances of both QPCs in linear response,
measured as a function of the respective gate voltage $V_3$, or
$V_8$. At low gate voltages, the QPCs are pinched-off and the
conductance is close to zero. With increasing gate voltage, 1D
channels successively open up~\cite{vanwees_wharam}. For both QPCs
we observe three conductance plateaus approximately quantized in
units of $G_0=2$e$^2$/h. With high bias
spectroscopy~\cite{kristensen} we find the spacing between the two
lowest subbands to be approximately 4~meV (3~meV) for the drive
(detector) QPC. The half-width of the energy window for opening a
1D subband is $\Delta\approx0.5$~meV in both QPCs.

\begin{figure}[t]\vspace{2mm}
\scalebox{0.92}{\includegraphics[clip]{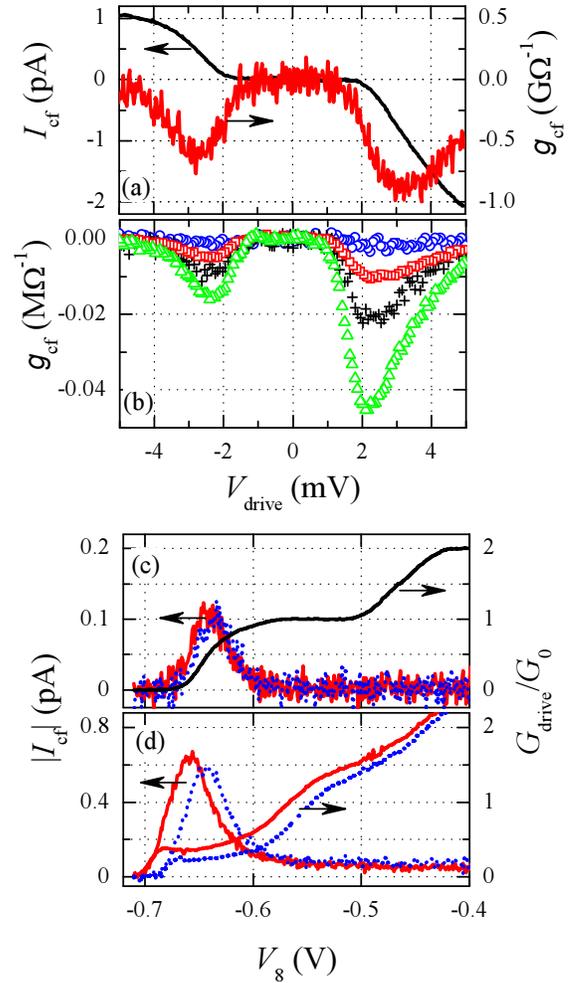}}\caption{(a) -
$I_\text{cf}$ and $g_\text{cf}$ for the nearly pinched-off
detector-QPC as a function of $V_\text{drive}$. (b) -
$g_\text{cf}$ measured for a set of $G_\text{det}$ values marked
by according symbols in Fig.~\ref{fig1}. (c,d) - Absolute value of
$I_\text{cf}$ as a function of the drive-QPC gate voltage $V_8$,
for $V_\text{drive}=\pm2.25~\text{mV}$ (c) and
$V_\text{drive}=\pm4~\text{mV}$(d). Also shown is the drive-QPC's
conductance in linear response (c) and its differential
conductance at $V_\text{drive}=\pm4~\text{mV}$ (d). Solid (dotted)
lines correspond to $V_\text{drive}<0$ ($>0$). In (a),(b) gates 7
and 9 are grounded, while in (c),(d) $V_7=V_9=-0.4$~V. The drive
bias modulation used to measure $g_\text{cf}$ is 92~$\mu$V~rms.
\label{fig2}}\vspace{-0.1in}
\end{figure}

Having characterized the QPCs, we turn to counterflow
measurements. Fig.~\ref{fig2}a shows the dc counterflow current,
$I_\text{cf}$, through the detector-QPC and the differential
counterflow conductance,
$g_\text{cf}\equiv~dI_\text{cf}/dV_\text{drive}$, as a function of
the bias on the drive-QPC. Here, the drive-QPC is tuned to nearly
half a conductance quantum $G_\text{drive}=G_0/2$, while the
detector-QPC is in the pinch-off regime (i.e. the lowest 1D
subband bottom is well above the Fermi level) with
$G_\text{det}\simeq10~\text{G}\Omega^{-1}$. Surprisingly, for
$|V_\text{drive}|\gtrsim~1$~mV, a finite current is observed in
the unbiased detector circuit. The direction of $I_\text{cf}$ is
opposite to that of the drive-QPC current $I_\text{drive}$. The dc
counterflow current is a threshold-like, nearly odd function of
$V_\text{drive}$. Correspondingly, the differential counterflow
conductance is negative and a nearly even function of
$V_\text{drive}$. The sign of $g_\text{cf}$ expresses a phase
shift of $\pi$ between the applied ac modulation of
$V_\text{drive}$ and the detected ac component of the counterflow
current.

Figures~\ref{fig2}c and~\ref{fig2}d show the absolute value of
$I_\text{cf}$ for the nearly pinched-off detector as a function of
the voltage on gate 8, which tunes the drive-QPC transmission. The
corresponding drive-QPC differential conductance curves are also
shown. For not too high $V_\text{drive}$ (Fig.~\ref{fig2}c), a
non-zero counterflow current is only detected in the region
between pinch-off and the first conductance plateau of the
drive-QPC. For higher $V_\text{drive}$ (Fig.~\ref{fig2}d)
$I_\text{cf}$ increases superlinearly with $V_\text{drive}$ at its
maximum and remains finite at higher gate voltages $V_8$. Since
the source bias effects the potential distribution near the
constriction, the nonlinear $1/2$ conductance plateau of the
drive-QPC shifts when changing $V_\text{drive}$~\cite{glazmann}.
This causes the shift of the extrema on Fig.~\ref{fig2}d as well
as the asymmetry of $g_\text{cf}$ in Fig.~\ref{fig2}a when
reversing the bias.

We proceed to study the counterflow effect in the regime of a more
opened detector-QPC. Figure~\ref{fig2}b plots
$g_\text{cf}$~\cite{remark1} as a function of $V_\text{drive}$ for
several values of $G_\text{det}$ between 0 and $G_0$ (marked with
the same symbols in Fig.~\ref{fig1}). The qualitative appearance
of $g_\text{cf}(V_\text{drive})$ is independent of $G_\text{det}$.
However, the amplitude of $g_\text{cf}$ is a strongly
non-monotonic function of the detector transmission. The
counterflow conductance reaches its maximum for
$G_\text{det}\approx G_0/2$ and decreases rapidly with further
increasing $G_\text{det}$. Note that the absolute value of
$g_\text{cf}$ is small, corresponding to a maximal ratio of the
counterflow and drive currents
$|I_\text{cf}/I_\text{drive}|\lesssim 10^{-3}$.

\begin{figure}[t]\vspace{2mm}
\scalebox{0.46}{\includegraphics[clip]{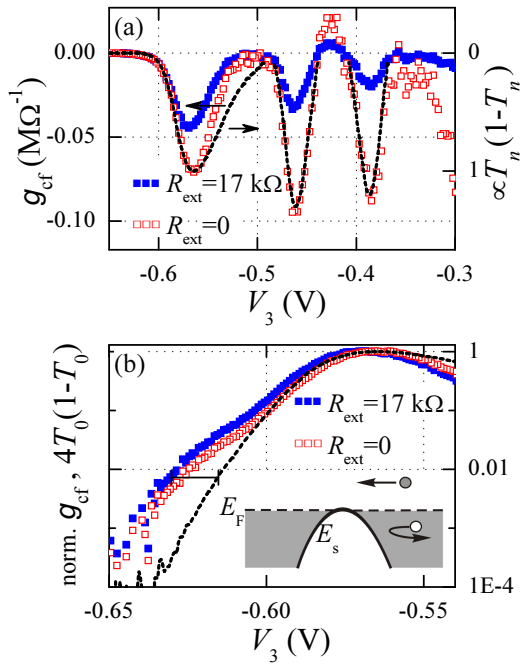}}\caption{
(a)- $g_\text{cf}$ as a function of the detector-QPC gate voltage
$V_3$. Filled symbols correspond to the $g_\text{cf}$ measured at
a finite external resistance $R_\text{ext}=17~{\rm k\Omega}$,
while open symbols show the corrected counterflow conductance
$R_\text{ext}=0$ (see text). Also shown are the transmission
functions $T_n(1-T_n)$ for the three lowest 1D subbands of the
detector-QPC (dashed lines), scaled to fit the corrected data.
During the $g_\text{cf}$ measurement the drive bias is modulated
with a 230~$\mu$V~rms signal about the mean value
$V_\text{drive}=+2.05$~mV. (b) - Normalized $g_\text{cf}$ (symbols
as in (a)) and transmission function of the lowest 1D detector-QPC
subband $4T_0(1-T_0)$ (dashed line) as a function of $V_3$. The
scale bar shows a gate voltage interval corresponding to a change
of the 1D subband energy by 0.5~meV. Inset: Sketch of possible
scattering processes of nonequilibrium electrons and holes at a
partially transmitting detector-QPC. \label{fig3}}\vspace{-0.1in}
\end{figure}
In Fig.~\ref{fig3}a $g_\text{cf}$ is plotted as a function of
$V_3$, controlling the detector transmission. $V_\text{drive}$ and
$V_8$ are adjusted for maximal $g_\text{cf}$ and kept fixed.
Confirming the trend seen in Fig.~\ref{fig2}b, the measured
$g_\text{cf}$ (solid symbols) strongly oscillates with increasing
$V_3$ and displays three pronounced maxima before the detector-QPC
is fully opened. The position of the {\it n}-th maximum ({\it n} =
0,1,2) is close to the value of $V_3$, where
$G_\text{det}/G_0\simeq~n+0.5$ (Fig.~\ref{fig1}). Here, the energy
$E_\text{S}^n$ of the bottom of the n-th 1D subband of the
detector-QPC aligns with the Fermi level of the leads
$E_\text{S}^{n}\simeq E_\text{F}$. In contrast, $g_\text{cf}$ is
close to zero for fully transmitting 1D channels
($G_\text{det}/G_0\simeq~n+1$). The overall magnitude of
$g_\text{cf}$ decreases with increasing $V_3$, hence
$G_\text{det}$. This is caused by a finite series resistance
$R_\text{ext}$ of the external circuit, which results in a
measured $g_\text{cf}$ lower than the case for an ideal
ammeter~\cite{remark2}. The corrected counterflow conductance,
$g^\text{ideal}_\text{cf}\equiv
g_\text{cf}\cdot(1+R_\text{ext}\cdot G_\text{det})$, corresponding
to $R_\text{ext}=0$, is shown in Fig.~\ref{fig3}a with open
symbols. The corrected maxima are roughly equal in size and
symmetric. Moreover, the shape of the {\it n}-th maximum compares
quite well with the corresponding function of the equilibrium
transmission $T_n(1-T_n)$, extracted from the detector conductance
data $T_n\equiv G_\text{det}/G_0-n$ (dashed lines in
Fig.~\ref{fig3}a).

In Figure~\ref{fig3}b we plot the normalized $g_\text{cf}$ and the
transmission function $4T_0(1-T_0)$ on a logarithmic scale near
the detector pinch-off. In the pinch-off regime (i.e. for
$T_0\ll1$) the transmission probability of a QPC is expressed as
${T_0(E)\propto {\rm
exp}([E-E_\text{S}^0]/\Delta)}$~\cite{glazmann}. Here $E$ is the
kinetic energy of current carrying electrons and $\Delta$ is the
half-width of the energy window for opening a 1D-subband. The
energy $E_\text{S}^0$ of the detector-QPC is controlled by gate 3
via $E_\text{S}^0\propto-|e|V_3$. This explains a nearly
exponential drop of the transmission function with decreasing
$V_3$ (Fig.~\ref{fig3}b). In contrast, the measured $g_\text{cf}$
drops considerably slower and remains finite even where the
detector-QPC is already pinched-off in equilibrium. This
experimentally observed excess contribution of the normalized
$g_\text{cf}$ versus $T_0(E_F)$ signals that the counterflow
current carrying electrons are excited well above the Fermi level.
Converting the shift in gate voltage (see the bar in
Fig.~\ref{fig3}b) to energy, we find a characteristic excitation
energy of $E^*\approx0.5$~meV. This is consistent with a recently
reported 1~meV bandwidth excitation provided by the drive-QPC for
electrons in a nearby double-dot quantum
ratchet~\cite{DQDratchet}.

\begin{figure}[t]\vspace{2mm}
\scalebox{0.92}{\includegraphics[clip]{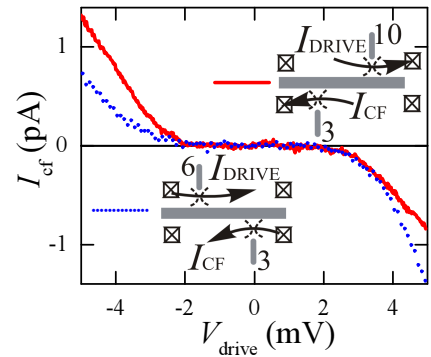}}\caption{Drive
bias dependence of the counterflow current through the pinched-off
detector-QPC for the drive-QPC formed with gate 6 (dotted line) or
gate 10 (solid line). The detector-QPC conductance is about
$G_\text{det}=5~\text{G}\Omega^{-1}$. The drive-QPCs are tuned to
provide the maximal effect. Insets: sketches of the two
counterflow measurements. The directions of currents are shown for
the case of $V_\text{drive}>0$. \label{fig4}}\vspace{-0.1in}
\end{figure}
Next we study the counterflow effect between spatially shifted
QPCs. Figure~\ref{fig4} shows $I_\text{cf}$ through the nearly
pinched-off detector-QPC as a function of the bias on the
drive-QPC, which is formed either with gate 10 or gate 6, while
gate 8 is now grounded (Fig.~\ref{fig1}). Despite the shift of the
drive-QPC position relative to the detector-QPC by about 300~nm,
the odd drive bias dependence of the counterflow current found in
Fig.~\ref{fig2} is practically preserved. This indicates that the
excitation of electrons in one of the leads of the detector-QPC is
not restricted to the close vicinity of the drive-QPC.

The oscillations of the counterflow conductance $g_\text{cf}$ in
Fig.~\ref{fig3} are reminiscent of thermopower oscillations that
have been investigated on individual QPCs~\cite{molenkamp,dzurak}.
This suggests that $I_\text{cf}$ is caused by an energetic
imbalance across the detector-QPC. If the bottom of the $n$-th
1D-subband of the detector-QPC is well separated from the
Fermi-energy in comparison to the characteristic excitation
energy, i.\,e$.$ if $|E_\text{S}^n-E_\text{F}|\gg E^*$, this
subband is either fully transmitting ($T_n(E)=1$) or closed
($T_n(E)=0$). In both cases electrons (holes) excited by $E^*$
above (below) $E_\text{F}$ are equally transmitted and
$g_\text{cf}=0$. In contrast, if $E_\text{S}^n\simeq E_\text{F}$
excited electrons are more likely transmitted than excited holes
(see inset of Fig.~\ref{fig3}b), resulting in $g_\text{cf}\neq0$.

The energetic imbalance across the detector-QPC we propose to be
caused by phonon-based energy transfer from the drive-QPC. The
excess energy of carriers injected across the drive-QPC is mainly
relaxed by emission of acoustic phonons. We consider the drive-QPC
in the non-linear regime near pinch-off where
$\mu_\text{S}-\mu_\text{D}\gg\Delta$ and the transmission
probability is strongly energy-dependent (the source and drain
leads are defined so that their chemical potentials satisfy
$\mu_\text{S}>\mu_\text{D}$). In this case electrons injected into
the drain lead have an excess energy of about
$e|V_\text{drive}|\equiv\mu_\text{S}-\mu_\text{D}$ whereas the
source lead remains essentially in thermal
equilibrium~\cite{heiblum}. Hence acoustic phonons are
predominantly generated in the drain lead of the drive-QPC.
Because of this asymmetry electron-hole pairs are excited
preferentially in the adjacent lead of the
detector-QPC~\cite{remark3}. This gives rise to $I_\text{cf}$
directed opposite to the current through the drive-QPC (and
$g_\text{cf}<0$). The data in Fig.~\ref{fig2} clearly show, that
the counterflow effect is only observed in the non-linear regime
of the drive-QPC.

For a rough estimate we consider injected electrons with a
momentum relaxation time of 60\,ps limited by elastic scattering
and an energy relaxation time of 1\,ns~\cite{Ridley,verevkin}.
Assuming isotropic phonon emission we estimate an energy transfer
ratio which can account for the observed value of
$I_\text{cf}/I_\text{drive}$ within one order of magnitude.

In summary, the current in a strongly biased drive-QPC generates a
current flowing in the opposite direction through an adjacent
unbiased detector-QPC. This counterflow current is maximal in
between the conductance plateaus of the detector-QPC. The effect
is most pronounced near pinch-off of the drive-QPC, where it
behaves strongly non-linear. We interpret the results in terms of
an asymmetric phonon-based energy transfer.

The authors are grateful to V.T.~Dolgopolov, A.W.~Holleitner,
C.~Strunk, F.~Wilhelm, I.~Favero, A.V.~Khaetskii,
N.M.~Chtchelkatchev, A.A.~Shashkin, D.V.~Shovkun and P.~H\"anggi
for valuable discussions and to D.~Schr$\ddot{\text{o}}$er and
M.~Kroner for technical help. We thank the DFG via SFB 631, the
BMBF via DIP-H.2.1, the Nanosystems Initiative Munich (NIM) and
VSK the A.~von~Humboldt foundation, RFBR, RAS, and the program
"The State Support of Leading Scientific Schools" for support.


\begin{thebibliography}{apssamp}
\bibitem{remark2} The input resistance
of the I-V amplifier, the ohmic contacts and wiring resistances
result in $R_\text{ext}=17$~k$\Omega$. The validity of the above
formula has been checked by applying an additional 47~k$\Omega$
resistor in series to $R_\text{ext}$.
\bibitem{remark3} $|I_\text{cf}|$ is reduced for
$V_\text{drive}<0$ ($>0$) and the drive-QPC shifted to the lh (rh)
side of the detector-QPC (Fig.~\ref{fig4}). This is understood in
terms of absorption of phonons in both leads of the detector-QPC.
\end{thebibliography}
\end{document}